\documentclass[prl,superscriptaddress,twocolumn]{revtex4-1}
\usepackage{amsmath}
\usepackage{amssymb}
\usepackage{bm}
\usepackage{epsfig}
\usepackage{graphicx}
\usepackage{color}
\usepackage[colorlinks]{hyperref}
\usepackage[english]{babel}
\usepackage{float}

\usepackage{empheq}
 
\usepackage[bbgreekl]{mathbbol}
\usepackage{units}

\begin{document}

\title{Nonlinear spectroscopy of high-spin fluctuations}

\author{A.~A.~Fomin}
\affiliation{Spin Optics Laboratory, St.\,Petersburg State University, 198504 St.\,Petersburg, Russia}

\author{M.~Yu.~Petrov}
\affiliation{Spin Optics Laboratory, St.\,Petersburg State University, 198504 St.\,Petersburg, Russia}

\author{I.~I.~Ryzhov}
\affiliation{Photonics Department, St.\,Petersburg State University, Peterhof, 198504 St.\,Petersburg, Russia}
\affiliation{Spin Optics Laboratory, St.\,Petersburg State University, 198504 St.\,Petersburg, Russia}

\author{G.~G.~Kozlov}
\affiliation{Spin Optics Laboratory, St.\,Petersburg State University, 198504 St.\,Petersburg, Russia}

\author{V.~S.~Zapasskii}
\affiliation{Spin Optics Laboratory, St.\,Petersburg State University, 198504 St.\,Petersburg, Russia}

\author{M.~M.~Glazov}
\affiliation{Ioffe Institute, 194021 St.\,Petersburg, Russia}
\affiliation{Spin Optics Laboratory, St.\,Petersburg State University, 198504 St.\,Petersburg, Russia}

\begin{abstract}

We investigate theoretically and experimentally fluctuations of high spin (\textit{S}\,$>$\,\nicefrac{1}{2}) beyond the linear response regime and demonstrate dramatic modifications of the spin noise spectra in the high power density probe field. Several effects related to an interplay of high spin and perturbation are predicted theoretically and revealed experimentally, including strong sensitivity of the spin noise spectra to the mutual orientation of the probe polarization plane and magnetic field direction, appearance of high harmonics of the Larmor frequency in the spin noise and the fine structure of the Larmor peaks. We demonstrate the ability of the spin-noise spectroscopy to access the nonlinear effects related to the renormalization of the spin states by strong electromagnetic fields.
\end{abstract}

\maketitle

\textit{Introduction}.\quad Spin manipulation by optical means including optical orientation effect, spin readout by Kerr and Faraday effects, inverse Faraday effect, coherent control, etc. are of high importance for the growing field of spintronics~\cite{RevModPhys.89.011004,PhysRevX.4.031027,Oka2019}. The spin-\nicefrac{1}{2} state serves as a prototype qubit 
two-level system for various applications, also being the simplest and most studied model for coherent control of quantum states~\cite{WolfScience01,glazov2018electron}. Regarding the quantum processing of information the use of high spins, i.e., \textit{S}\,$>$\,\nicefrac{1}{2}, offers serious advantages due to a higher number of controlled degrees of freedom~\cite{PhysRevA.67.012310,Lanyon:2008aa,NisbetJones2013,PhysRevA.97.022115,Soltamov:2019aa}. While the fine structure of the spin states can be tuned by a magnetic field, the non-magnetic spin control is one of the challenging tasks offering significant advantages. One example of such kind phenomena is the spin-orbit interaction which couples spin and orbital degrees of freedom and, consequently, allows one to control the electron spins electrically~\cite{dyakonov_book,Rashba03}. Another is an optical quantum state control, which was demonstrated to be possible not only for trapped cold atoms, but even for strongly interacting systems~\cite{Mitchell:2020}. 

Another possibility is to use the nonlinear interactions of spin systems with electromagnetic fields. Tailored periodic driving of quantum systems, also known as Floquet engineering, provides vast playground for developing the systems with novel properties and applications in various areas of physics including ultrafast spectroscopy and high-speed spintronics.
 In particular, due to the inverse Faraday effect~\cite{PhysRev.143.574,pitaevskii_inv:farad}, the circularly polarized light field produces effective, optically induced, magnetic field acting on the spins~\cite{Ryzhov:2016aa}. For a two-level spin-\nicefrac{1}{2} system any perturbation can be reduced to an effective magnetic field~\cite{allen1975optical}.

The situation becomes very rich for the \textit{S}\,$>$\,\nicefrac{1}{2}. Firstly, there are higher-order spin arrangements which are not reduced to the spin polarization, e.g., the spin alignment where the average spin orientation is absent, but the spins are predominantly aligned parallel or antiparallel to a certain axis (for more details on this effect, see the precedent work of our group~\cite{PhysRevResearch.2.012008}). Accordingly, one may seek the ways to control high spins not only by the circularly polarized, but also by a linearly polarized light. 

In this paper we investigate, both theoretically and experimentally, the spin fluctuations in the atomic gas in the presence of strong linearly polarized probe beam. The spin noise spectroscopy has initially emerged as virtually perturbation-free technique to study the spin dynamics and fine structure of the spin states~\cite{aleksandrov81,PhysRevLett.80.3487,Mitsui:2000nx,Crooker_Noise,Zapasskii:13,Oestreich:rev,Glazov:15}. Increasing the probe beam intensity makes it possible to address a plethora of non-equilibrium phenomena including incoherent effects related with relaxation processes and also the coherent phenomena~\cite{PhysRevA.84.043851,PhysRevLett.113.156601,glazov:sns:jetp16,PhysRevB.95.241408,PhysRevB.97.081403}. Here we demonstrate that the renormalization of the spin \textit{S}\,$>$\,\nicefrac{1}{2} states by the electromagnetic field causes a number of non-linear effects, which include: the light-induced fine splittings of the spin states, induced anisotropy of the spin system, i.\,e., the dependence of the spin noise spectra on the orientation of the probe beam with respect to the magnetic field, and appearance of high harmonics of the Larmor precession frequency in the spin noise. We illustrate experimentally theoretical predictions taking Cs atomic vapour as testbed, because in this case the multiplet corresponding to the total spin $S={3,4}$ is easily accessible by the spin noise spectroscopy~\cite{PhysRevA.97.032502,PhysRevResearch.2.012008}. The results presented in the work can open up a way for system sublevels optical control, which can be executed by combination with radiofrequency addressing of the states~\cite{Poshakinsky2019} or `active' spin noise spectroscopy~\cite{Sharipova2019}.

\textit{Theory.}\label{sec:phen}\quad Let us consider a spin \textit{S}\,$>$\,\nicefrac{1}{2} system, for example, an atom in the presence of the intense monochromatic laser beam (probe). We assume that the frequency of the laser~$\omega$ exceeds by far all characteristic frequencies of the spin system, particularly, the Larmor frequency, but can be close to one of the optical resonances. We assume, however, that the probe beam is not perfectly resonant with any of transitions and neglect in in this work possible absorption of the light related to real transitions between atomic states. Such situation is typical in the spin noise spectroscopy~\cite{aleksandrov81,PhysRevResearch.2.012008}.

We are interested in the effects of the intense probe beam on the spin noise spectrum. Since the beam is off-resonant and the absorption is disregarded, only virtual transitions are possible and the description of the probe beam effects can be reduced to an effective Floquet Hamiltonian which describes low-energy dynamics of the spin system, while the energy scales of $\hbar\omega$ and higher are integrated out. 
The general form of the effective Hamiltonian can be determined from the symmetry arguments. The effective (dressed) Hamiltonian for the spin $\bm F$ in the presence of the external magnetic field $\bm B$ and the probe beam with the electric field $\bm E$ in the form 
\begin{equation}
\bm E = \bm E_0 e^{-\mathrm i \omega t} + {\rm c.c.},
\end{equation}
with $\bm E_0$ being the complex amplitude, in quadratic in the $|E_0|$ approximation reads
\begin{equation}
\label{H:eff}
\mathcal H = \gamma (\bm B \cdot \bm F) + \mathrm i \mathcal B ([\bm E_0 \times \bm E^*_0] \cdot \bm F) + \mathcal C \{(\bm E_0\cdot \bm F)(\bm E^*_0 \cdot \bm F)\}_s.
\end{equation}
Here $\mathcal B$ and $\mathcal C$ are the parameters, $\gamma$ is the gyromagnetic ratio, the non-linear in $\bm B$ contributions to the Hamiltonian~\eqref{H:eff} are disregarded. The first term here is the Zeeman effect caused by the external field. The derivation of the remaining terms is based on the method of invariants in the group theory and relies on the fact that the Hamiltonian is invariant under the point-group transformations of the system~\cite{birpikus_eng}. In our case, this is the group of all transformations of the three-dimensional space, i.e., $SO(3)\times I$: the transformations include all the rotations, inversion and their compositions. Thus, to derive Eq.~\eqref{H:eff} we need to build the invariant combinations of the spin components and probe field. We restrict the consideration by bi-linear in $\bm E_0$, $\bm E_0^*$ effects, i.e., the contributions to the Hamiltonian linear in the light intensity. Corresponding field combinations transform either as a pseudovector, $[\bm E_0\times \bm E_0^*]$ (i.e., asymmetric second-rank tensor which is dual to a pseudovector), and as the symmetric second-rank tensor, $\{E_{0,\alpha}E_{0,\beta}^*\}_s$. The former ones couple with the pseudovector $\bm F$ and provide an invariant as $[\bm E_0\times E_0^*]\cdot \bm F$  [second term in Eq.~\eqref{H:eff}], while the latter ones couple with symmetrized combinations $\{F_{\alpha}F_{\beta}^*\}_s$ [third term in the Hamiltonian~\eqref{H:eff}].
The term $\mathrm i \mathcal B ([\bm E \times \bm E^*] \cdot {\bm F})$ describes the inverse Faraday effect~\cite{pitaevskii_inv:farad,PhysRev.143.574,Ryzhov:2016aa}: The circularly polarized probe induces the effective Zeeman splitting of the atomic levels. Indeed, the product $\mathrm i [\bm E \times \bm E^*]$ transforms as a pseudovector and describes the helicity of the photon, thus it couples with spin just like an external magnetic field. The term $ \mathcal C \{(\bm E_0\cdot \bm F)(\bm E^*_0 \cdot \bm F)\}_s$ with curly brackets standing for symmetrization of the operators describes the renormalizations of the energies due to the linearly polarized light. Note that while individual factors $(\bm E_0\cdot \bm F)$ and $(\bm E_0^*\cdot \bm F)$ are pseudoscalars, their product is scalar. For spin $S=\nicefrac{1}{2}$ such symmetrized product is unimportant, since the products of spin-\nicefrac{1}{2} components are reduced to the first powers of spins. Thus, the term~$\propto \mathcal C$ appears for $S>\nicefrac{1}{2}$ only, it is somewhat similar to the quadrupole splittings of the nuclear states by the strain or electric field gradients~\cite{glazov2018electron}.

The parameters $\mathcal B$ and $\mathcal C$ can be calculated in the framework of the time-dependent perturbation theory~\cite{cohen-atom-photon} and can be estimated as $\mathcal B, \mathcal C \sim |d|^2/\Delta = \varpi_R^2/(|E_0|^2\Delta)$, where $d$ is the dipole matrix element of the optical transition, $\varpi_R$ is the Rabi frequency, and $\Delta=\omega_0 - \omega$ is the probe detuning from the optical resonance $\omega_0$~\footnote{The detailed microscopic theory will be presented elsewhere.}.

Equation~\eqref{H:eff} demonstrates the key effects of the intense probe beam. To begin with, let us briefly consider circularly polarized radiation where the symmetric combinations $E_{0,\alpha}E_{0,\beta}^*+E_{0,\alpha}^*E_{0,\beta}$ vanish and 
\begin{equation}
\label{circular}
\mathrm i [\bm E_0 \times \bm E_0^*] = P_c \bm n |E_0|^2,
\end{equation} 
where $\bm n$ is the unit vector along the light propagation direction, $P_c$ is the degree of circular polarization. Thus, the second term in Eq.~\eqref{H:eff} can be presented in the Zeeman form with the effective -- optical -- magnetic field 
\begin{equation}
\label{optical:field}
\bm B_{\rm eff} = \gamma^{-1} P_c \mathcal B \bm n.
\end{equation} This field combines with the external field resulting in a shift of the Larmor frequency and corresponding modifications of the spin noise spectra. The effect of ``optical'' magnetic field created by elliptically polarized light was studied in detail theoretically and demonstrated experimentally on a semiconductor system in a preceding work~\cite{Ryzhov:2016aa} of our group.

\begin{figure}[bt]
\includegraphics[width=\linewidth]{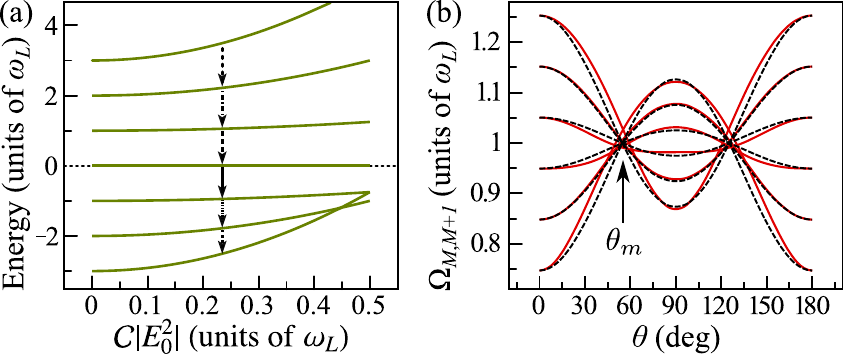}
\caption{(a) Energy spectrum of the dressed Hamiltonian~\eqref{H:eff} {for $S=3$} at the fixed magnetic field for various probe beam intensities and $\theta=0$. Arrows indicate transition frequencies. (b) Transition frequencies $\Omega_{M,M+1}$ calculated after Eq.~\eqref{transition:1} (black dashed lines) and by numerical diagonalization of the Hamiltonian~\eqref{H:eff} (red solid lines) at $\mathcal C|E_0|^2/\omega_L=1/20$. Arrow indicates the magic angle $\theta_m$.}\label{fig:spec}
\end{figure}

\begin{figure*}[t]
\includegraphics[width=\linewidth]{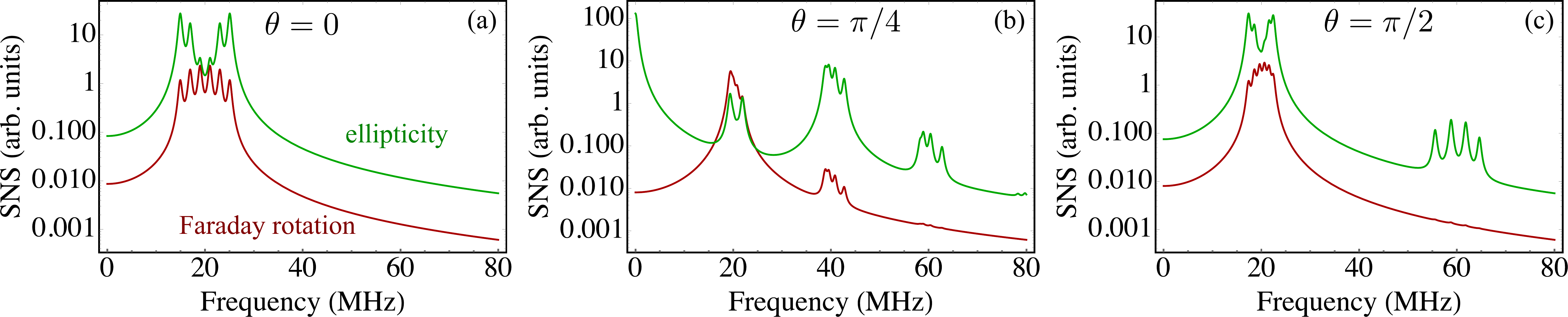}
\caption{Noise spectra of Faraday rotation $\mathcal N_r(\nu)$ and ellipticity $\mathcal N_e(\nu)$ calculated using 
	 Eqs.~\eqref{noise:gen} and \eqref{noise:Sz}. Panels (a), (b), and (c) correspond to $\theta=0$, $\pi/4$, and $\pi/2$, respectively. The parameters of calculation are: $\omega_L=20$~MHz, $\mathcal C|E_0|^2/\omega_L=1/20$, and $\Gamma=0.4$~MHz.}\label{fig:SNS}
\end{figure*}

In what follows we focus on the effects of linearly polarized probe where the circular polarization vanishes, $P_c=0$, and the Hamiltonian along with the Zeeman contribution of the external field contains also the quadratic in spin $\bm F$ terms. Let $\theta = \angle \bm B, \bm E_0$ be the angle between the magnetic field and the polarization vector of the probe. To illustrate the effect, we assume that 
\begin{equation}
\label{cond:0:weak}
\mathcal C |E_0|^2 \ll |\gamma \bm B|.
\end{equation}
Thus, the third term in the Hamiltonian~\eqref{H:eff} can be considered as a small perturbation. Accordingly, the eigenstates are characterized by a projection $M=-S,{-S+1},\ldots,S-1,S$ of $\bm F$ onto the $\bm B$-axis. Making use of the first-order perturbation theory, we evaluate the matrix elements of the last term in Eq.~\eqref{H:eff} on the basic functions with the definite spin projection $M$ on the magnetic field (eigenfunctions of the first term in the Hamiltonian) we evaluate $\mathcal E_M$. As a result, the differences between the eigenenergies $\mathcal E_M$ or transition frequencies acquire the form
\begin{multline}
\label{transition}
\Omega_{MM'} \equiv \frac{\mathcal E_{M'} -\mathcal E_M}{\hbar} 
= \omega_L (M'-M) + \\
+\frac{1}{2} \mathcal C |E_0|^2\left(2\cos^2{\theta} - \sin^2{\theta}\right) (M'^2-M^2).
\end{multline}
Here 
\begin{equation}
\label{Larmor}
\omega_L = \gamma B,
\end{equation}
is the Larmor spin precession frequency in the magnetic field $\bm B$. Thus, due to the probe-induced effect, the spectrum of the spin in the magnetic field is no longer equidistant and the transition frequencies are no longer multiples of the $\omega_L$, but demonstrate a fine structure determined by the second term in Eq.~\eqref{transition}, as schematically illustrated in Fig.~\ref{fig:spec}(a). The fine structure depends on the orientation of the probe beam polarization plane with respect to the magnetic field and described, in accordance with the symmetry requirements, by the second angular harmonics of the angle $\theta$. Interestingly, the fine structure in the quadratic in the field approximation vanishes at a ``magic'' angle
\begin{equation}
\label{magic}
\theta_m = \arctan{{\sqrt{2}}}.
\end{equation}
The numerical diagonalization of the Hamiltonian~\eqref{H:eff} demonstrates that the suppression of splittings takes place only in the $|E_0|^2$ order, compare solid and dashed curves in Fig.~\ref{fig:spec}(b).

These features in the energy spectrum are clearly revealed in the spectrum of spin fluctuations. Using the effective Hamiltonian we can formally define the fluctuation spectrum of any observable $O$ as 
\begin{equation}
\label{noise:gen}
(O^2)_\nu = \frac{2\Gamma\hbar^2}{2S+1}\sum_{ij} \frac{|O_{ij}|^2}{(\hbar\nu - \mathcal E_j + \mathcal E_i)^2+\hbar^2\Gamma^2},
\end{equation}
where $i$ and $j$ enumerate the eigenstates of the system, $\Gamma$ is the phenomenological broadening. For example, in the geometry where the magnetic field $\bm B \parallel x$, for the spin-$z$ component fluctuations we have
\begin{multline}
\label{noise:Sz}
(F_z^2)_\nu = \frac{\Gamma}{2(2S+1)}\sum_{M=-S}^{S-1} \frac{(S+M+1)(S-M)}{(\nu - \Omega_{M,M+1})^2+\Gamma^2}
\\
+\{\nu\to -\nu\},
\end{multline}
Therefore, the intense probe beam results in the fine structure of the spin fluctuation spectrum: Instead of a single peak at the Larmor frequency the system would be characterized with a series of $2F$ peaks with the frequencies [cf. Eq.~\eqref{transition}]:
\begin{equation}
\label{transition:1}
\Omega_{M} \equiv \Omega_{M,M+1} = \omega_L +\frac{1}{2} \mathcal C |E_0|^2\left(2\cos^2{\theta} - \sin^2{\theta}\right) (2M+1).
\end{equation}

\begin{figure}
\includegraphics[width=\linewidth]{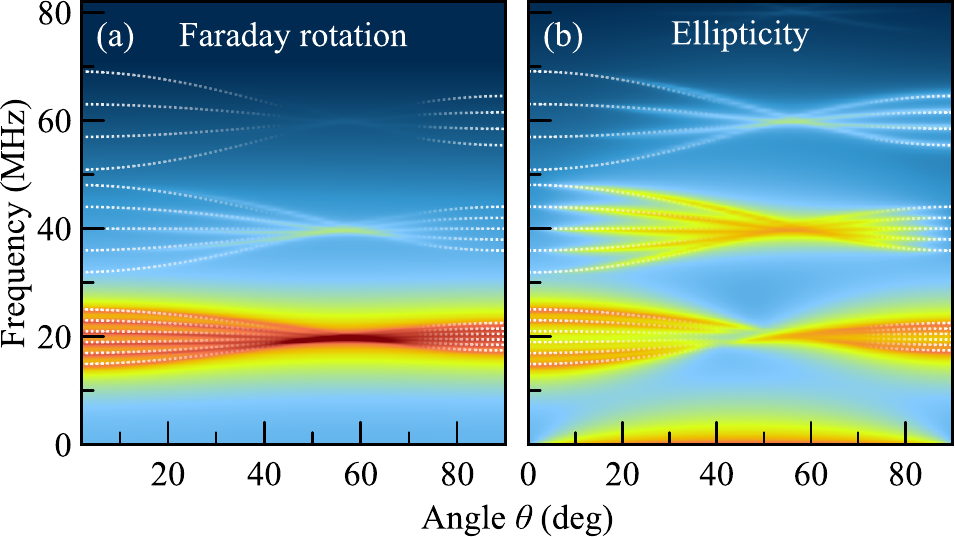}
\caption{Calculated noise spectra of Faraday rotation $\mathcal N_r(\nu)$ (a) and ellipticity $\mathcal N_e(\nu)$ (b) as a function of probe azimuth $\theta$ and frequency $\nu$. The transition frequencies calculated after Eq.~\eqref{transition:1} are shown in white dotted lines, which are made transparent in the vicinity of magic angle not to hinder the spectral features). The parameters of calculation are: $\omega_L=20$~MHz, $\mathcal C|E_0|^2/\omega_L=1/20$, and $\Gamma=0.4$~MHz.}
\label{fig:SNS:dens}
\end{figure}

The fluctuations of the spin $z$-component are revealed by the fluctuations of the Faraday rotation angle of the probe beam passing through the atomic vapor in the transparency region~\cite{aleksandrov81,glazov2018electron,PhysRevResearch.2.012008}
\begin{subequations}
\label{noise:all}
\begin{equation}
\label{rot:noise}
\mathcal N_r(\nu) \propto (F_z^2)_\nu.
\end{equation}
If, instead of the Faraday rotation, the fluctuations of ellipticity are detected, then, for \textit{S}\,$>$\,\nicefrac{1}{2} the noise of the spin alignment
\begin{multline}
\label{ell:noise}
\mathcal N_e(\nu) \propto \sin^2{2\theta}\left[(F_x^2-F_y^2)^2\right]_\nu + 2\cos^2{2\theta}\left(\{F_xF_y\}_s^2\right)_\nu
\end{multline}
is detected, see Refs.~\cite{PhysRevResearch.2.012008,kozlov2020polarimetric} for details.
\end{subequations}
The fluctuation spectra of the Faraday rotation and ellipticity calculated after general Eqs.~\eqref{noise:gen}, \eqref{noise:Sz}, and \eqref{noise:all} are demonstrated in Fig.~\ref{fig:SNS}. Figure~\ref{fig:SNS:dens} shows the same spectra as functions in the form of density plots. The positions of peaks are well reproduced by analytical formula~\eqref{transition:1}, as indicated by the dashed lines in Fig.~\ref{fig:SNS:dens}.

It follows from Eq.~\eqref{H:eff} that, due to the term $\mathcal C \{(\bm E_0\cdot \bm F)(\bm E^*_0 \cdot \bm F)\}_s$, the states with different components $M$ are mixed at $\theta \ne 0, \pi$. It gives rise to the higher harmonics of the Larmor frequency in the spin noise spectra, particularly, it gives rise to the $2\omega_L$ and $3\omega_L$ peaks, see Fig.~\ref{fig:SNS}(b,c) and Fig.~\ref{fig:SNS:dens}. The higher-order in $E_0$ terms (omitted in the Hamiltonian~\eqref{H:eff}) give rise to the higher harmonics.

To conclude, let us summarize the main effects of the intense probe beam.
\begin{itemize}
\item Circularly polarized beam gives rise to an effective optical magnetic field, Eq.~\eqref{optical:field} (see~\cite{Ryzhov:2016aa} for details).
\item Linearly polarized probe beam results in the fine structure of the transition energies grouped in the vicinity of the Larmor frequency, Eq.~\eqref{transition}. 
\item Accordingly, it results in the fine structure of the Larmor peak in the spin noise spectrum.
\item The effective energy spectrum and, correspondingly, the spin noise spectrum depends on the mutual orientation of the probe beam polarization and the magnetic field.
\item The higher Larmor harmonics in the spin noise spectra appear as a result of the symmetry breaking by the intense probe beam.
\end{itemize}
It should be noted that in high-precision applications (for example, for accurate measurements of the Earth magnetic field) one should take into account not only the described above features, but also other effects affecting the shape of the EPR spectral lines, in particular, the nonlinearity of the Zeeman effect and the accompanying heading-error effect~\cite{Budker2018}. In the framework of this work, we omit these fine effects due to the fact that under our conditions the width of the spin noise line significantly exceeded the value of the splittings caused by the nonlinear Zeeman effect. The linewidth in our case was determined not only by the field inhomogeneity, but also by the optically induced line broadening which origin will be discussed elsewhere.
\begin{figure}[t]
	\includegraphics[width=0.8\linewidth]{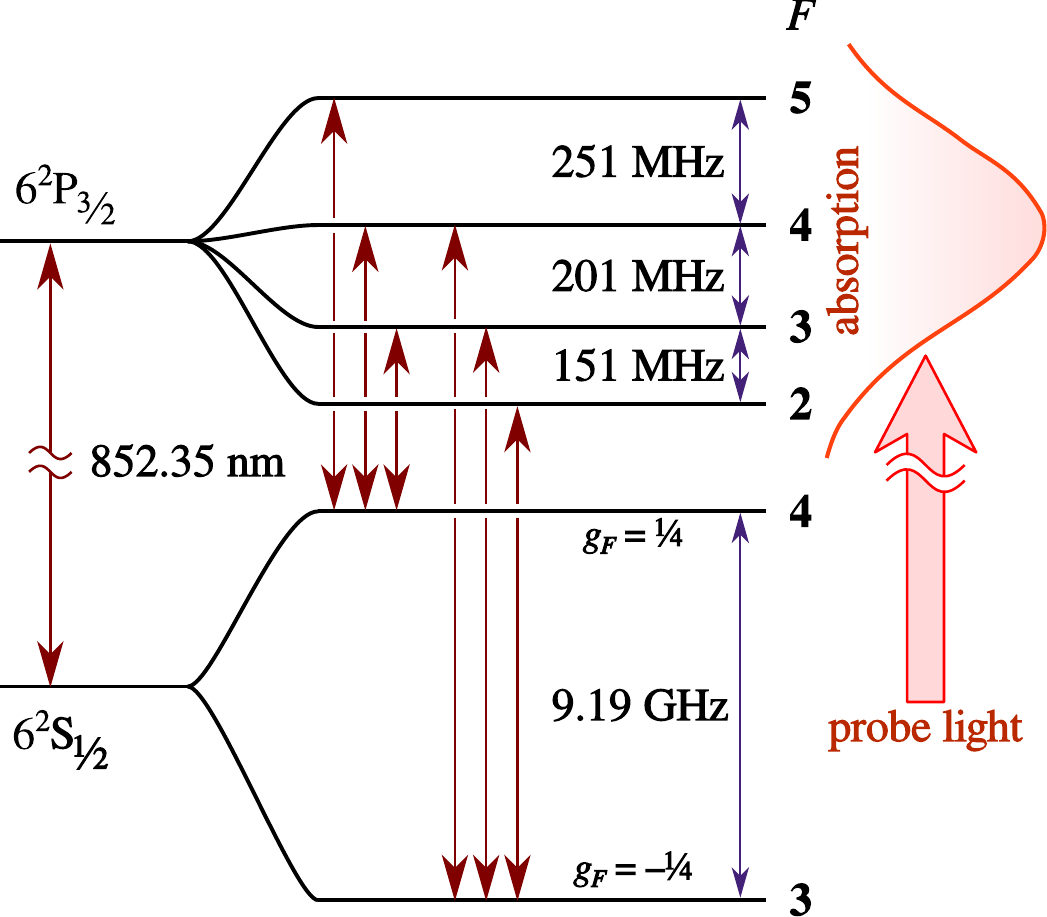}
	\caption{Energy-level diagram of Cesium atom in the range of D2 line. The spectral position of the probe light, which was slightly detuned from the center of D2 optical transition, is shown schematically.}
	\label{fig:fig4}
\end{figure}

\textit{Experimental illustrations.}\label{sec:exp}\quad As a high-spin system we used cesium atom with its ground state comprised of two hyperfine (hf) components with total angular momenta ${S} = 3$ and ${S} = 4$. Energy level diagram of Cs atom is shown schematically in Fig.~\ref{fig:fig4}. The measurements were performed at long-wavelength component of the D2 line corresponding to the transition $^2S_{\nicefrac{1}{2}}$ (${S} = 4$)$\:\to\!{}^2P_{\nicefrac{3}{2}}$ ({$S = 3,4,5$}). Because of Doppler broadening, the hf structure of the excited state is not resolved, and, in the absorption spectrum of the D2 line, one can observe only two component spaced by hf splitting of the grounds state (9.19 GHz). As a laser source we used a ring Ti:sapphire laser tuned in resonance with the long-wavelength wing of the absorption line (as shown in Fig.~\ref{fig:fig4}). The cylindrical cell $20 \times 20$~mm in size with cesium vapor at a temperature of 60\ldots70$^\circ$C, was mounted inside a pair of the Helmholtz-type coils, creating magnetic field of around 1.6 mT aligned across the laser light propagation. 

\begin{figure*}[t]
\centering
\includegraphics[width=\linewidth]{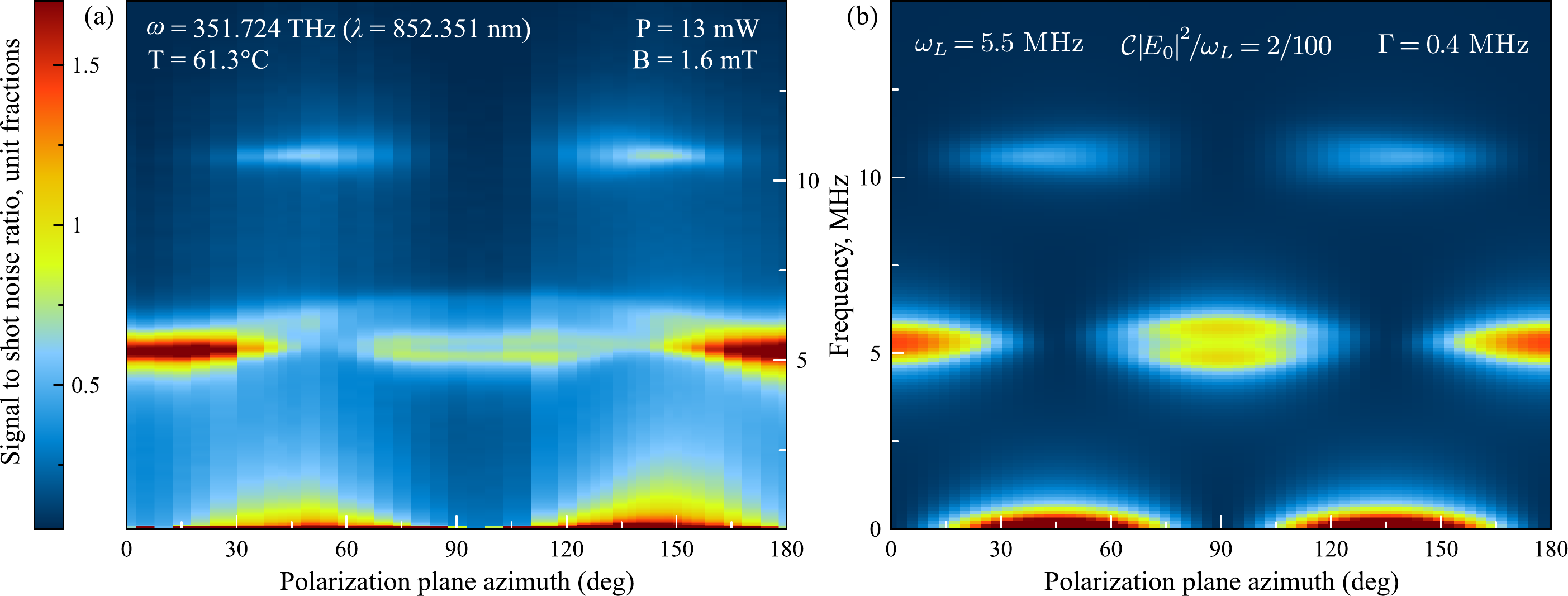}
\caption{Azimuthal dependence of the ellipticity-noise spectra of Cs atoms, under conditions of resonant probing: experimental data (a) and model (b). The experimental conditions and the parameters of calculation are shown at the top of panels (a) and (b), respectively.}
\label{fig:fig5}
\end{figure*}

Design of the experimental setup was common for spin noise spectra measurement. Briefly, the linearly polarized collimated laser beam, 4 mm in diameter, after passing through the cell, was detected by a standard differential polarimetric photoreceiver, with its output signal processed in real time by a broadband fast Fourier transform radio-frequency spectrum analyzer Tektronix RSA5103A. The light power density on the sample was 0.08 W/cm$^2$ for the total laser beam intensity of 10 mW. Placing a quarter-wave plate after the cell with one of its principal axes aligned along the light beam polarization allowed us to detect fluctuations of the light ellipticity, rather than the noise of the Faraday rotation. The polarization plane azimuth of the incident beam was controlled by rotation of a half-wave plate installed before the sample. All the measurements were performed with the polarimetric photoreceiver set to the balance.

Figure~\ref{fig:fig5}(a) shows the measured ellipticity noise spectrum vs the angle between the light polarization plane and magnetic field direction. To compare these results with predictions of our theoretical treatment, we have to take into account that conditions of our measurements were far from the idealized assumptions of the theory. Specifically, our experiments did not meet the strict requirements of sufficiently large detuning accepted in the theory: In the experiment a combination of $\mathcal N_e(\nu)$ and $\mathcal N_r(\nu)$, as well as real processes of the probe beam absorption may affect the results of the measurements. What is, perhaps, more important is that, because of inhomogeneity of the light power density over the probe beam cross-section, the fine structure of the spectra could not be as pronounced as in theory. So, to make our calculations more realistic, we enhanced the relative broadening of the spin-noise spectrum, $\Gamma/\omega_L\approx 0.07$. Under these conditions, the $\Gamma$ exceeded the scale of the light-induced splittings, $\Gamma/(\mathcal C|E_0|^2) \approx 3.5$. 

The results of these calculations are presented in Fig.~\ref{fig:fig5}(b). The general pattern of this angular dependence perfectly agrees with the experimental data: at $\theta = 0^\circ$ and $90^\circ$, the splitting is the greatest while at some intermediate angle this splitting tends to vanish. Within the accuracy of our measurements, position of the latter point well correlates with predictions of the theory. At the double Larmor frequency, we can see the peak of the spin-alignment noise with its azimuthal dependence being in agreement with the model.

To conclude, we have demonstrated both theoretically and experimentally that for \textit{S}\,$>$\,\nicefrac{1}{2} the interaction of spin with electromagnetic field results in renormalization of its energy spectrum. Particularly, the linearly polarized intense probe beam results in the appearance of the fine structure of the spin levels which strongly depends on the mutual orientation of the light polarization plane and the magnetic field. The effect can be straightforwardly detected in the spin noise spectroscopy where the fine structure of Larmor precession peak in the transverse magnetic field is revealed along with higher harmonics of the Larmor precession frequency. The developed analytical theory is illustrated by experimental data on Cs vapors. The uncovered phenomena open up the possibilities to control the spin states non-magnetically via the Floquet engineering of the Hamiltonian. One should 

\begin{acknowledgments}
\textit{Acknowledgements.}\quad Theoretical work of M.\,M.\,G. was partially supported by Saint-Petersburg State University, research Grant No. 51125686. The work was fulfilled using the equipment of the SPbU resource center `Nanophotonics'. Experimental investigations were funded by RFBR grant No.~19-52-12054 which is highly appreciated. I.\,I.\,R. acknowledges the support of experimental work by Presidential Grant No.~MK-2070.2018.2.
\end{acknowledgments}

\end{document}